\documentstyle[epsf]{elsart}
\begin{document}
\begin{frontmatter}
%
%%%%%%%%%%%%%%%%%%%%%%%%%%%%%%%%%%%%%%%%%%%%%%%%%%%%%%%%%%%%%
\title{Interfaces and Grain Boundaries of Lamellar Phases}
%%%%%%%%%%%%%%%%%%%%%%%%%%%%%%%%%%%%%%%%%%%%%%%%%%%%%%%%%%%%%
%
\author[tau]{Simon Villain-Guillot},
\author[mpi]{Roland R. Netz},
\author[tau]{David Andelman} and
\author[seattle]{Michael Schick}
\address[tau]{School of Physics and Astronomy, Tel Aviv University,
Tel Aviv 69978, Israel}
\address[mpi]{Max-Planck-Institut f\"ur Kolloid und
Grenzfl\"achenforschung\\Kantstr. 55, 14513 Teltow-Seehof, Germany}
\address[seattle]{Department of Physics, Box 351560, University of
 Washington\\Seattle WA 98195-1560, USA}
%
%%%%%%%%%%%%%%%%%%%%%%%%%%%%%%%%%%%%%%%%%%%%%%%%%%%%%%%%%%%%%
\begin{abstract}
%%%%%%%%%%%%%%%%%%%%%%%%%%%%%%%%%%%%%%%%%%%%%%%%%%%%%%%%%%%%%
%
Interfaces between lamellar and disordered phases, and grain boundaries
within lamellar phases, are investigated employing a simple Landau free
energy functional. The former are examined using analytic, approximate
methods in the weak segregation limit, leading to density
profiles which can extend over many wavelengths of the lamellar phase. The
latter are studied numerically and exactly. We find a change from smooth
chevron configurations typical of small tilt angles to distorted omega
configurations at large tilt angles in agreement with experiment.
\end{abstract}
\begin{keyword}
Lamellar Phases, Modulated Phases, Interfaces, Grain Boundaries,
Block copolymers
\PACS{61.72.M, 82.65.D, 64.60.Cn} 
\end{keyword}
\end{frontmatter}
%%%%%%%%%%%%%%%%%%%%%%%%%%%%%%%%%%%%%%%%%%%%%%%%%%%%%%%%%%%%%
\section{Introduction}
%%%%%%%%%%%%%%%%%%%%%%%%%%%%%%%%%%%%%%%%%%%%%%%%%%%%%%%%%%%%%

  Lamellar phases are found in many systems, such as ferrofluids,
   mixtures of lipids and water, and
  melts of 
  diblock copolymers \cite{science}. 
  Whenever such phases occur, one expects
  to observe grain boundaries between phases of different
  orientations. While such boundaries have been the subject of much
  study in solids, there has been very little work on
  their occurrence in complex fluids. The recent experimental work of Gido and 
  Thomas \cite{thomas} and Hashimoto et al. \cite{hashimoto} on 
  grain boundaries in diblock copolymer
  systems showed that the conformation of the boundary was a strong
  function of the angle between grains. Whereas for small angles, the
  lamellae varied smoothly from one orientation to the other, (a ``chevron''
  configuration), for larger
  angles the lamellae became quite distorted, sending out a piece of
  lamellae nearly parallel to the boundary
  itself (an ``omega'' configuration). Finally, when the interface
is parallel to the lamellae of one of the two adjoining phases,
the lamellae  became disjunct, abutting one
  another in a ``T-junction''. In this paper, a simple Landau free
  energy is shown to produce precisely this behavior. 

  In addition, we consider the interface between coexisting lamellar and
disordered phases. Again such interfaces occur in many systems: those of
pure diblock copolymer melts \cite{batesschulz}, of mixtures of
homopolymer and diblock \cite{batesmaurer}, of lipids and
water \cite{briggs}, of small amphiphiles, oil and water \cite{strey} etc.
The interface is studied analytically in the weak-segregation limit {\it
i.e.} in which the ordering of the lamellae can be described by a single
Fourier amplitude.

%%%%%%%%%%%%%%%%%%%%%%%%%%%%%%%%%%%%%%%%%%%%%%%%%%%%%%%%%%%%%
\section{The Model}
%%%%%%%%%%%%%%%%%%%%%%%%%%%%%%%%%%
We consider a three-dimensional system in which 
the ordering can be described by a scalar
order parameter $\phi$, and employ the dimensionless Ginzburg-Landau 
free energy
functional (rescaled by $k_B T$)
\begin{eqnarray}
{F}[\phi]&=&\int\left\{-\frac{\chi}{2}\phi^2
+\frac{1}{2}(\nabla^2\phi)^2-\frac{1}{2}(\nabla\phi)^2\right.
\nonumber\\
&& \left.
+~~\frac{1-\phi}{2}\ln\frac{1-\phi}{2}+\frac{1+\phi}{2}\ln\frac{1+\phi}{2}
\right\}{\rm d}V
\label{model}
\end{eqnarray}
The interaction term, of strength $\chi$, induces the
system to order ($\phi\neq 0$) 
as $\chi$ increases.  That the
coefficient of the gradient squared term is negative expresses the
system tendency to make this ordered phase a modulated one in which
the order parameter varies in space. 
The positive coefficient of the Laplacian squared term ensures
that the spatial variation does not become too large. 
Finally,
the logarithmic entropy of mixing terms 
oppose the tendency to order, preferring a
state in which $\phi$ vanishes. 
%Note that we have used the freedom to 
%establish the scale of length, of
%energy, and that of
Note that 
the order parameter $\phi$ is limited  to have magnitude less than
or equal to unity. The coefficients
 of the Laplacian
squared and gradient squared terms 
are chosen to be $\pm 1/2$, respectively, setting the 
length scale in the problem.
For convenience, we first turn to the
weak segregation limit and the interface between lamellar and disordered
phases.

%%%%%%%%%%%%%%%%%%%%%%%%%%%%%%%%%%%%%%%%%%%%%%%%%%%%%%%%%%%%%
\section{The Lamellar-Disorder Interface in Weak Segregation}
%%%%%%%%%%%%%%%%%%%%%%%%%%%%%%%%%%%%%%%%%%%%%%%%%%%%%%%%%%%%%%

We assume here that the ordering is weak, {\it i.e.} that $|\phi|$ is
small. In that case the free energy functional can be expanded to fourth
order in $\phi$. We restrict ourselves to lamellar phases, and introduce
the Fourier representation 
$\phi(x)=\sum\phi_{ k}\exp(i{k}{x})$, with $\phi_k=\phi_{-k}^*$ as the
order parameter is real. In this representation 
the free energy per unit volume is
\begin{equation}
\frac{{F}}{V}=\frac{1}{2}\sum_k A_k\phi_k\phi_{-k}+
\frac{1}{12}\sum_{k,p,q}\phi_k\phi_p\phi_q\phi_{-(k+p+q)} 
%-\ln 2
\end{equation}
where a non relevant constant term is omitted from $F$, and 
$A_k\equiv [1-\chi-k^2+k^4]$.
The $k$-mode which becomes critical at the highest temperature is that for
which $A_k$ is minimal. This occurs at
$k=1/\sqrt{2}\equiv q$, at which $A_{q}=3/4 -\chi$. When
$\phi_{q}$ is small, all other $\phi_k$, save $\phi_0$, can be ignored as they are
proportional to integer powers of $\phi_{q}$.
Hence
\begin{equation}
\label{free}
\frac{{F}}{V}\approx
  \frac{1}{2}(1-\chi)\phi_0^2+\frac{1}{12}\phi_0^4+
\left(\frac{3}{4}-\chi+\phi_0^2\right)\phi_{q}\phi_{-q}
+\frac{1}{2}(\phi_q\phi_{-q})^2.
%-\ln 2
\end{equation}

Minimizing this free energy with respect to 
$\phi_q\phi_{-q}=|\phi_q|^2$, one obtains
\begin{equation}
\label{op}
\phi_q\phi_{-q}|_{min}=(\chi-\phi_0^2-3/4).
\end{equation}
As this must be positive, it follows that in
the lamellar phase,\ $0\leq \phi_0^2\leq \chi-3/4$. A line of continuous
transitions from the disordered to lamellar phase occurs when $\phi_q$
vanishes, {\it i.e.} along the line
\begin {equation}
\chi_c=\phi_0^2+3/4.
\end{equation}
Upon substitution of the minimum value of 
the order parameter, Eq. (\ref{op}), into the free energy of
Eq. ({\ref{free}), one finds that in the lamellar phase,
\begin{equation}
\label{freelam}
\frac{{F}^L(\phi_0, \chi)}{V}=-\frac{1}{2}\left(\frac{3}{4}-\chi\right)^2
+\frac{1}{2}\left(\chi-\frac{1}{2}\right)\phi_0^2-\frac{5}{12}\phi_0^4,
\end{equation}
as compared to the free energy of the disordered phase which, from Eq. 
(\ref{free}) with $\phi_q=0$, is
\begin{equation}
\label{freedis}
\frac{{
    F}^D(\phi_0,\chi)}{V}=\frac{1}{2}(1-\chi)\phi_0^2+\frac{1}{12}\phi_0^4.
\end{equation}
The line of continuous transitions ends at a tricritical point at which 
the free energy of the lamellar phase, Eq. (\ref{freelam}), is no longer
convex with respect to $\phi_0$. This occurs at the point
$\phi_0^t=1/4$, and $\chi^t=13/16$. For larger values of $\chi$ and
$\phi_0$, 
the system undergoes phase separation, with lamellar and disordered phases 
coexisting. The value of $\phi_0$ in each, denoted $\phi_0^L$ and
$\phi_0^D$, respectively, is obtained from the conditions of the equality
of chemical potentials, $\mu$,
\begin{equation}
\frac{1}{V}\frac{\partial {F}^L(\phi_0^L,\chi)}{\partial \phi_0^L}=
\frac{1}{V}\frac{\partial {F}^D(\phi_0^D,\chi)}{\partial \phi_0^D}
\end{equation}
and of grand potentials $G/V\equiv F/V-\mu \phi_0$
\begin{equation}
F^L(\phi_0^L,\chi)-\frac{\partial {F}^L(\phi_0^L,\chi)}{\partial
  \phi_0^L}\phi_0^L=F^D(\phi_0^D,\chi)-
\frac{\partial {F}^D(\phi_0^D,\chi)}{\partial \phi_0^D}\phi_0^D.
\end{equation}
Once $\phi_0^L$ is known, the amplitude of the modulation in the
coexisting lamellar phase follows from Eq. (\ref{op}).
We now turn to the calculation of the interfacial profile, $\phi(x)$,
between these phases.

The interfacial free energy per unit area 
(and in units of $k_B T$) is the minimum of the functional
\begin{equation}
\sigma[\phi]=\int {\rm d}x\left\{\frac{1}{V}F[\phi]-
\mu\phi(x)
-\frac{1}{2V}(G^L +G^D)\right\}.
\end{equation}
where the first term $F[\phi]/V$ is just the local free energy
density as appearing under the spatial integration of 
Eq. (\ref{model}), and $G^L$ and $G^D$ are the lamellar and disordered
bulk grand potentials, respectively, as defined above.
The Euler-Lagrange equation which results from this minimization is to
be solved for the profile subject to the boundary conditions 
$\phi(x) \rightarrow \phi_0^D$ as $x\rightarrow \infty$, and
$\phi(x)\rightarrow \phi_0^L+2|\phi_q|\cos(qx)$ as $x\rightarrow -\infty$.
Rather than solve this equation directly, we use the {\em ansatz}
$\phi(x)=\phi_0^D[1+f(x)]/2+\phi_0^L[1-f(x)]/2+g(x)|\phi_q|\cos(qx)$
where 
$f(x)$ and
$g(x)$ approach 1 and 0, respectively as $x\rightarrow \infty$, and 
$-1$ and 2, respectively as $x\rightarrow -\infty$. 
Near the
tricritical point, we can expand the amplitudes $\phi_0^D$, $\phi_0^L$,
and $\phi_q$ about the values they take at the tricritical point, 1/4,
1/4, and 0, respectively, in a power series in $\chi-\chi^t$. Solving the
equations for $f(x)$ and $g(x)$ to lowest order in this small parameter,
we find \cite{simon}
\begin{equation}
g(x)=\frac{2{\rm e}^{-U}}{\sqrt{1+{\rm e}^{-2U}}},
\end{equation}
\begin{equation}
f(x)=\tanh(U)-\frac{18}{5\cosh^2(U)}(\chi-\chi^t)
\end{equation}
where $U(x)=2(3/5)^{1/2}(\chi-\chi^t)x.$
A profile for the case $\chi=0.9$ is shown in Fig. 1. Several lamellae
participate in the interfacial region for this weak first order transition. 
\begin{figure}
%fig1
\epsfysize=16\baselineskip
\centerline{\vbox{\epsffile{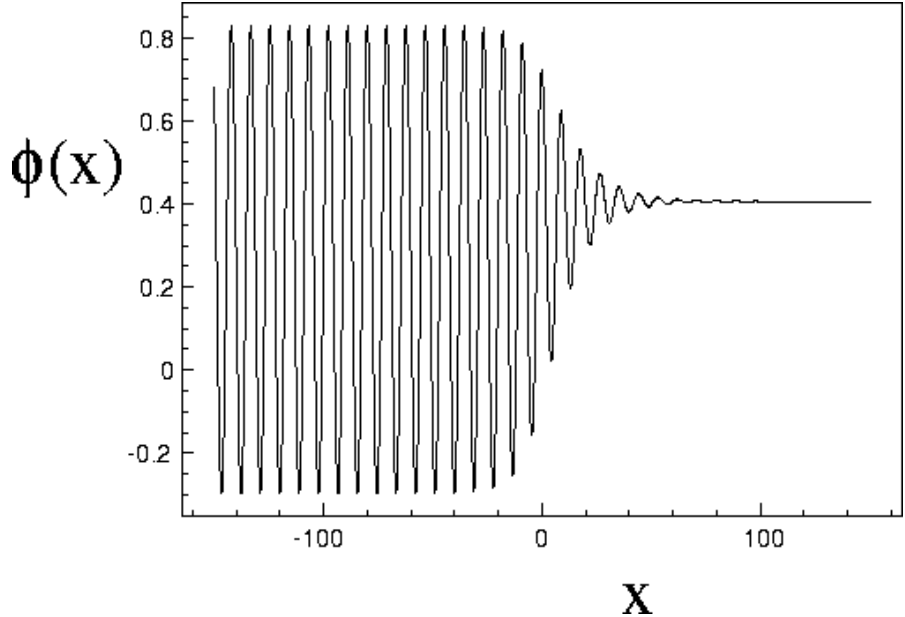}}}
\caption{Interfacial profile between lamellar and disordered phases 
for $\chi=0.9$ (weak segregation limit).}
\end{figure}
As the
tricritical point is approached, the width of the interface, governed by
the bulk correlation length, $\xi$, will diverge, of course. In our
calculation, we find the expected mean-field scaling \cite{widom}
$\xi\sim1/(\chi-\chi^t)$. The wavelength characterizing the lamellar
phase, however, remains at $2\pi/q$. Thus there are many oscillations
of the order parameter within the interface as the tricritical point is
approached.  The evaluation of the interfacial tension itself proceeds
in a similar manner, and one obtains the scaling $\sigma\sim
(\chi-\chi^t)^2$ expected within mean-field theory 
of tricritical points \cite{widom}.

We close this section by noting that 
these interfacial phenomena associated with the occurrence of
a tricritical point can be observed for uniaxial lamellar systems, such
as thermotropic liquid crystal at the nematic/smectic-A  transition
\cite{Litster}.
If phases of hexagonal symmetry are
allowed, the line of critical 
phase transitions between disordered and
lamellar phases, which ends at the tricritical point,
 is pre-empted by a line of first-order 
phase transition between
disordered and hexagonal phases. Only the continuous transition at
$\phi_0=0$ remains. Thus, the interface we have considered 
close to a tricritical point exists only between
metastable phases. Nonetheless, the general results we obtain should
apply also to isotropic systems
when  there is a weak first-order
transition between disordered and lamellar phases. As noted earlier,
this is the case in
many systems: pure diblock copolymers \cite{batesschulz}, copolymer
mixtures \cite{batesmaurer}, lipid and water mixtures \cite{briggs},
and ternary mixtures of small amphiphiles, oil and water \cite{strey}. 
In each case,
one expects an interface in which the transition from the lamellar to
the disordered phase occurs over a length scale, $\xi$, which is quite
large with respect to the wavelength $\lambda$ of the lamellar phase
itself. The interfacial tension between phases, $\sigma$, will be small
if the first-order transition is weak.

%%%%%%%%%%%%%%%%%%%%%%%%%%%%%%%%%%%%%%%%%%%%%%%%%%%%%%%%%%%%%
\section{Grain Boundaries in Lamellar Phases of Complex Systems}
%%%%%%%%%%%%%%%%%%%%%%%
 We next consider grain boundaries between lamellar phases and do so by
considering the free energy of Eq. (\ref{model}). In contrast to the
analytic, approximate treatment above, in this section we minimize
the free energy numerically, and {\em exactly} \cite{netz}. 
To do so, we discretize
space into a 200 by 200 lattice, and minimize the free energy functional
with respect to the order parameter amplitude on the 40,000 lattice
sites by means of a conjugate-gradient method.  Appropriate boundary conditions
are imposed to bring about a grain boundary. We set the parameter
$\chi$ and $\phi_0$ such that the lamellar phase is stable. In the
results below, $\chi=1$, and $\phi_0=0$.  Fig. 2(a) shows a symmetric
tilt boundary with an angle of $28.08^{\circ}$ between the lamellae
normals.
\begin{figure}
%fig2a,2b,2c
\epsfxsize=0.33\linewidth
\centerline{\vbox{\epsffile{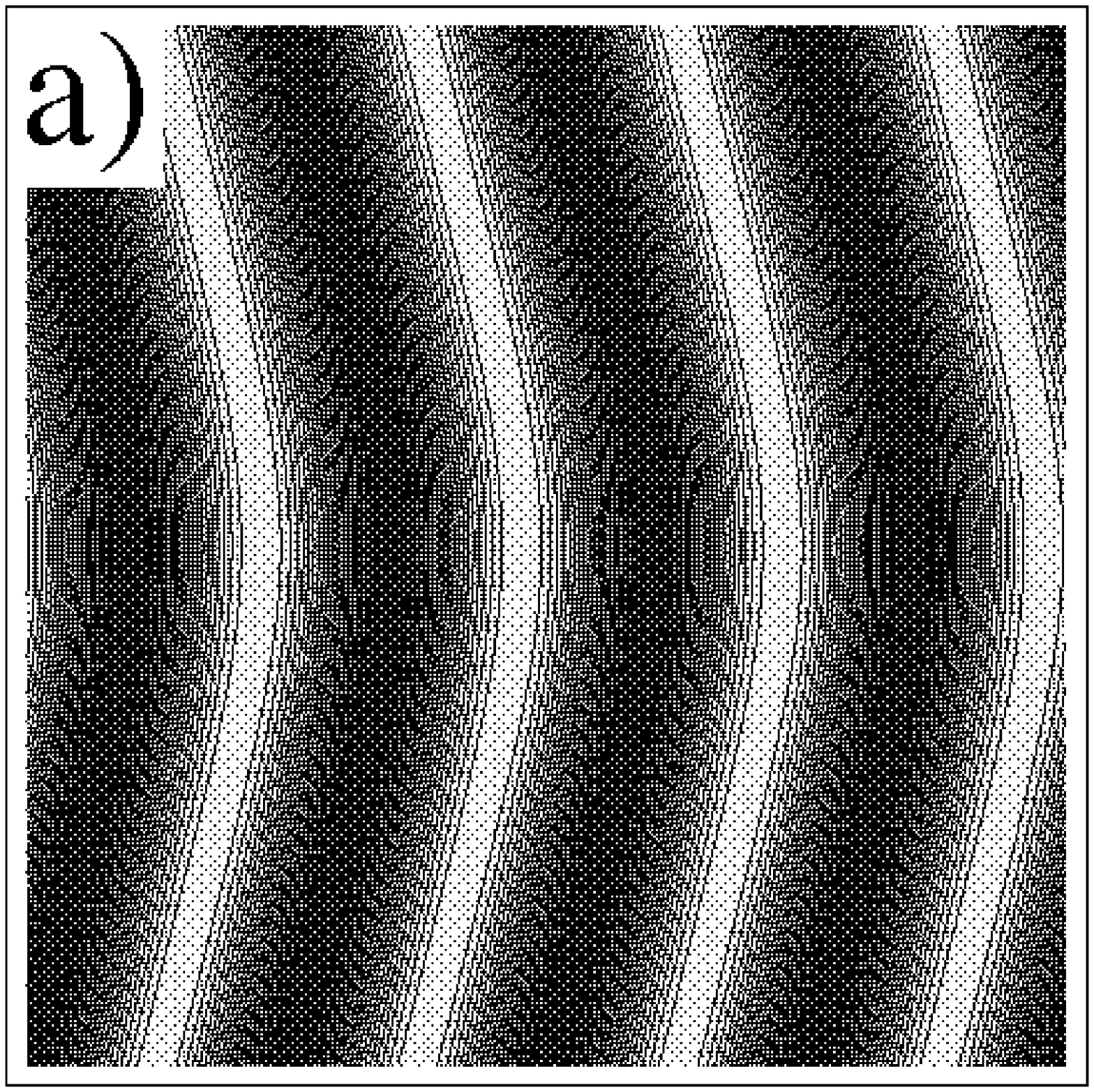}}
\epsfxsize=0.33\linewidth
            \vbox{\epsffile{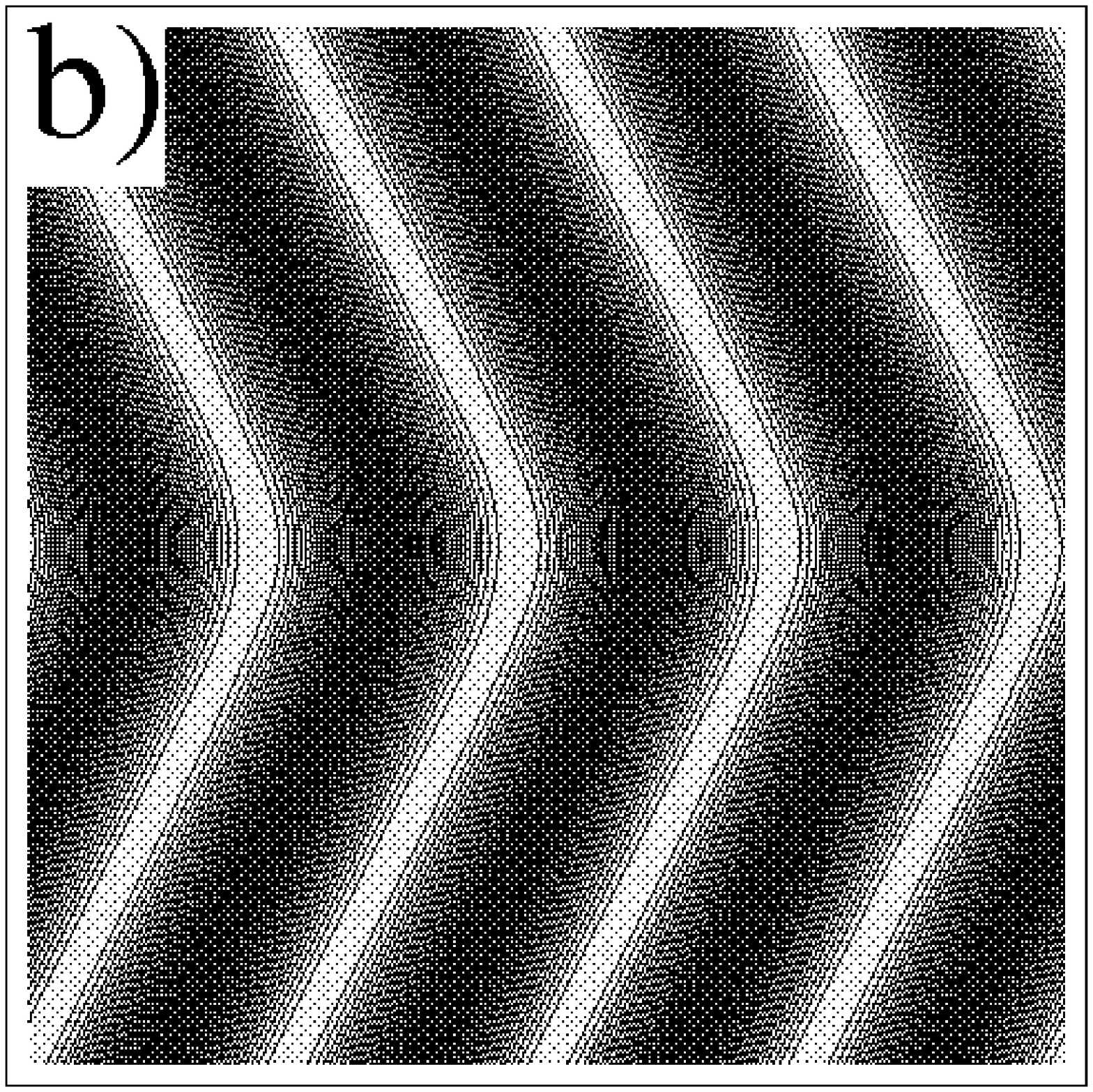}}
\epsfxsize=0.33\linewidth
            \vbox{\epsffile{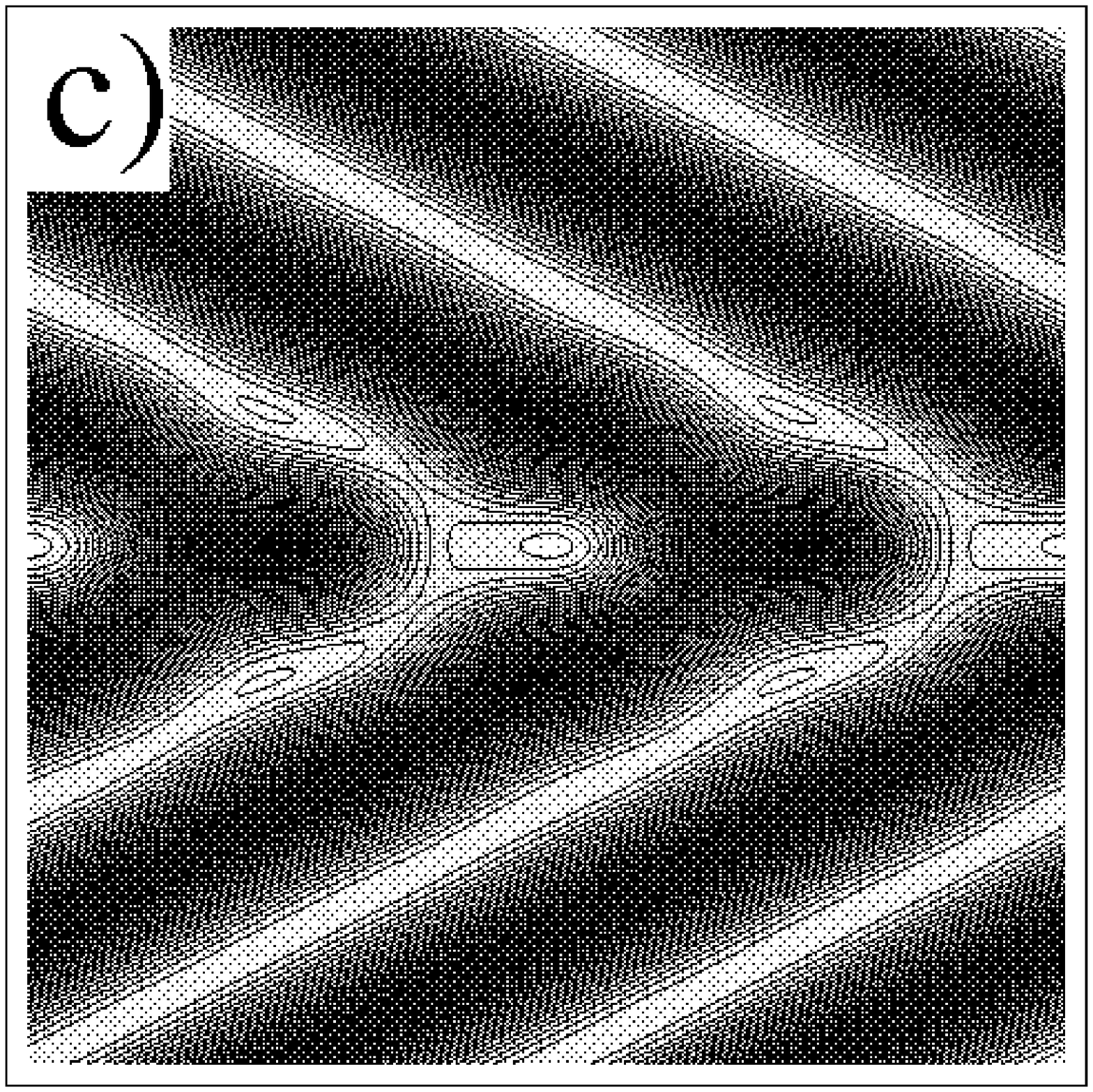}}}
\caption{Order parameter profiles for symmetric 
tilt grain boundaries
  for $\phi_0=0$ and $\chi=1$. The order parameter ranges from $-1$ to
  1, and is represented by 20 grayscales. (a) $\theta=28.08^{\circ}$;
(b) $\theta=53.14^{\circ}$, (c) $\theta=126.86^{\circ}$.}
\end{figure}
 The configuration is clearly that denoted ``chevron'' in
ref. \cite{thomas}. This smooth configuration remains when the tilt
angle is increased, in Fig. 2(b), to $\theta=53.14^{\circ}$. 
However when the angle is
increased to 126.86$^{\circ}$, as in Fig. 2(c), the configuration has
changed markedly to that denoted ``omega'' in ref. \cite{thomas}. The
change in configuration is clearly due to the fact that the spacing
between lamellae {\em at} the grain boundary itself,
$\lambda/\cos(\theta/2)$, is so much larger than the preferred spacing,
$\lambda$. By sending out the tip observed in Fig. 2(c), the distance
between regions of the same sign of the order parameter is reduced.

As the system is brought close to the transition to the disordered
phase, we observe a pronounced reconstruction of the grain boundary in
terms of a square-like modulation. This is shown in Fig. 3 for
$\chi=0.78$. 
\begin{figure}
%fig3
\epsfysize=16\baselineskip
\centerline{\vbox{\epsffile{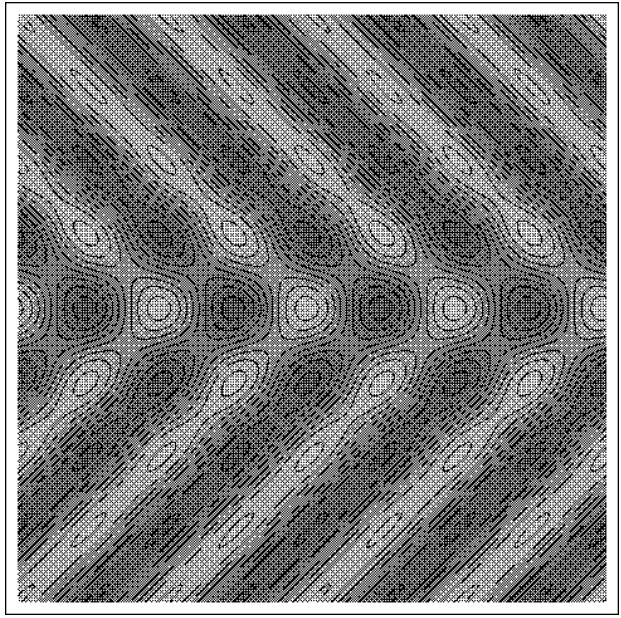}}}
\caption{Symmetric tilt grain boundary with $\theta=90^{\circ}$, and 
 $\chi=0.78.$}
\end{figure}
The symmetric tilt-boundary free energy per unit area,
$\sigma_{TB}$, decreases as the transition is approached and, for
$\phi_0=0$, vanishes as $\sigma_{TB}\sim(\chi-3/4)^{3/2}$ in accord with
mean-field predictions. As the angle of the tilt-boundary approaches
zero, its free energy vanishes as $\theta^3$; as the angle approaches
$\pi$, the energy is expected to vanish linearly with $(\pi-\theta)$ in
accord with a description in terms of independent dislocations of finite
creation energy.

Finally we turn to the ``asymmetric" T-junction. In Fig. 4, we show
results for this junction, again at $\chi=1$ and $\phi_0=0$.
In contrast to the chevron and omega configurations, only one of the two
types of domains is continuous across the interface in the T-junction.
\begin{figure}
%fig 4
\epsfysize=16\baselineskip
\centerline{\vbox{\epsffile{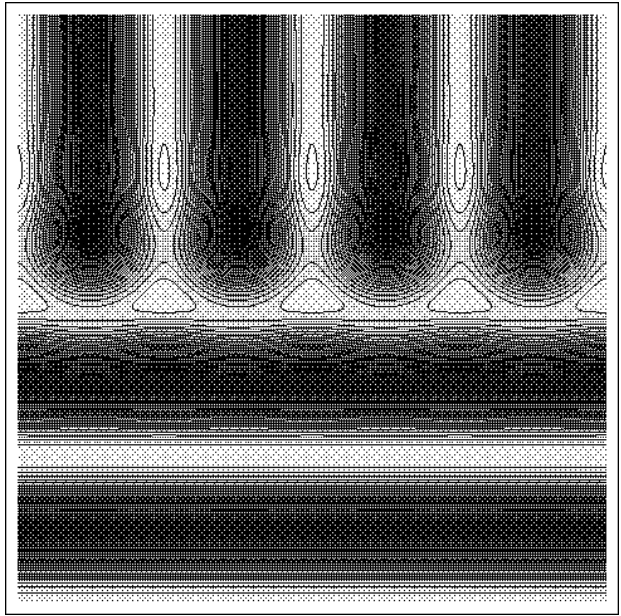}}}
\caption{Asymmetric grain boundary (T-junction) 
between two perpendicular 
  lamellar phases at
 $\phi_0=0$ and $\chi=1$.}
\end{figure}
 Note the
enlarged endcaps of the terminated lamellae, a feature which is also
clearly observable in experiment. We find that these endcaps become less
prominent as $\chi$ is increased. The free energy of this junction also
vanishes as $(\chi-3/4)^{3/2}$ as expected in mean-field theory. More
interestingly, we find that the free energy of this boundary is much
less than that brought about by inserting one wavelength of disordered
phase between the grains. Thus, the reconstruction at the boundary is a
far more efficient way of making the transition from one orientation to
another than that of a grain boundary melting. 

In summary, we have employed a simple Ginzburg-Landau free energy and calculated
analytically, and approximately, the properties of an interface between
a lamellar phase and a disordered one in a weak segregation limit. We
have used the same free energy to calculate numerically, and exactly,
the form of the grain boundaries  between lamellar phases. The agreement
with experiment is excellent. As this has been obtained from a Landau
free energy, this phenomena must result from a full
mean-field  calculation as well \cite{matsen}. 
Finally, we are able to make additional
predictions concerning the reconstruction of these interfaces as the 
temperature of the system is varied. In physical systems (block 
copolymers and others), phases of hexagonal and cubic symmetry are 
found in addition to the lamellar ones. It will be
most interesting to investigate the
interfaces and grain boundaries 
of these phases, as we have done here for the lamellar one.

We acknowledge support from the United States-Israel Binational Science
Foundation under grant no. 94-00291, and the National Science Foundation
under grant no. DMR 9531161. One of us (SVG) thanks the French Ministry
of Foreign Affairs for a research fellowship.

%%%%%%%%%%%%%%%%%%%%%%%%%%%%%%%%%%%%%%%%%%%%%%%%%%%%%%%%%%%%%

%\section*{Figure Captions}
%
%\noindent{\bf Fig. 1}~~~
%{Interfacial profile between lamellar and disordered phases
%for $\chi=0.9$ (weak segregation limit).}
%
%
%\bigskip
%\noindent{\bf Fig. 2}~~~
%{Order parameter profiles for symmetric tilt grain boundaries for 
%$\phi_0=0$ and $\chi=1$. The order parameter ranges from $-1$ to 1, and 
%is represented by 20 grayscales. (a) $\theta=28.08^{\circ}$;
%(b) $\theta=53.14^{\circ}$, (c) $\theta=126.86^{\circ}$.}
%
%\bigskip
%\noindent{\bf Fig. 3}~~~
%{Symmetric tilt grain boundary with $\theta=90^{\circ}$, and $\chi=0.78.$}
%
%\bigskip
%\noindent{\bf Fig. 4}~~~
%{Asymmetric grain boundary (T-junction) between two perpendicular
%lamellar phases at $\phi_0=0$ and $\chi=1$.}
%
%%%%%%%%%%%%%%%%%%%%%%%%%%%%%%%%%%%%%%%%%%%%%%%%%%%%%%%%%%%%%
%\end{multicols}
\end{document}